# Theory of citing

M.V. Simkin and V.P. Roychowdhury

Department of Electrical Engineering, University of California, Los Angeles, CA 90095-1594

We present empirical data on misprints in citations to twelve high-profile papers. The great majority of misprints are identical to misprints in articles that earlier cited the same paper. The distribution of the numbers of misprint repetitions follows a power law. We develop a stochastic model of the citation process, which explains these findings and shows that about 70-90% of scientific citations are copied from the lists of references used in other papers. Citation copying can explain not only why some misprints become popular, but also why some papers become highly cited. We show that a model where a scientist picks few random papers, cites them, and copies a fraction of their references accounts quantitatively for empirically observed distribution of citations.

## I. Statistics of misprints in citations

"Now let us come to those references to authors, which other books have, and you want for yours. The remedy for this is very simple: You have only to look out for some book that quotes them all, from A to Z …, and then insert the very same alphabet in your book, and though the imposition may be plain to see, because you have so little need to borrow from them, that is no matter; there will probably be some simple enough to believe that you have made use of them all in this plain, artless story of yours. At any rate, if it answers no other purpose, this long catalogue of authors will serve to give a surprising look of authority to your book. Besides, no one will trouble himself to verify whether you have followed them or whether you have not, being no way concerned in it…"

<div style="text-align:right">Miguel de Cervantes, Don Quixote</div>

When scientists are writing their scientific articles, they often use the method described in the above quote. They can do this and get away with it until one day they copy a citation, which carries in it a DNA of someone else's misprint. In such case, they can be identified and brought to justice, similar to how biological DNA evidence helps to convict criminals, who committed more serious offences than that.

Our initial report [1] led to a lively discussion[1] on whether copying a citation is a proof of not reading the original paper. Alternative explanations are worth exploring; however, such hypotheses should be supported by data and not by anecdotal claims. It is indeed most natural to assume that a copying citer also failed to read the paper in question (albeit this cannot be rigorously proved). *Entities must not be multiplied beyond necessity.* Having thus shaved the critique with Occam's razor, we will proceed to use the term non-reader to describe a citer who copies.

As misprints in citations are not too frequent, only celebrated papers provide enough statistics to work with. Let us have a look at the distribution of misprints in citations to one renowned paper (No. 5 in Table I.1), which at the time of our initial inquiry [1], that is in late 2002, had accumulated 4,301 citations. Out of these citations 196 contained misprints, out of which only 45 were distinct. The most popular misprint in a page number appeared 78 times.

**Table I.1.** Papers, misprints in citing which we studied.

| No | Reference |
|---|---|
| 1 | K. G. Wilson, Phys. Rev. 179, 1499 (1969) |
| 2 | K. G. Wilson, Phys. Rev. B 4, 3174 (1971) |
| 3 | K. G. Wilson, Phys. Rev. B 4, 3184 (1971) |
| 4 | K. G. Wilson, Phys. Rev. D 10, 2445 (1974) |
| 5 | J.M. Kosterlitz and D.J. Thouless, J. Phys. C 6, 1181 (1973) |
| 6 | J.M. Kosterlitz, J. Phys. C 7, 1046 (1974) |
| 7 | M.J. Feigenbaum, J. Stat. Phys. 19, 25 (1978) |
| 8 | M.J. Feigenbaum, J. Stat. Phys. 21, 669 (1979) |
| 9 | P. Bak, J. von Boehm, Phys. Rev. B 21, 5297 (1980) |
| 10 | P. Bak, C. Tang, and K. Wiesenfeld, Phys. Rev. Lett. 59, 381 (1987) |
| 11 | P. Bak, C. Tang, and K. Wiesenfeld, Phys. Rev. A 38, 364 (1988) |
| 12 | P. Bak and C. Tang, J. Geophys. Res. B 94, 15635 (1989) |

As a preliminary attempt, one can estimate the ratio of the number of readers to the number of citers, *R*, as the ratio of the number of **distinct** misprints, *D*, to the **total number** of misprints, *T*. Clearly, among *T* citers, $T - D$ copied, because they repeated someone else's misprint. For the *D* others, with the information at hand, we have no evidence that they did not read, so according to the presumed innocent principle, we assume that they did. Then in our sample, we have *D* readers and *T* citers, which lead to:

$$R \approx D/T. \tag{I.1}$$

---

[1] See, for example, the discussion "Scientists Don't Read the Papers They Cite" on Slashdot:
http://science.slashdot.org/article.pl?sid=02/12/14/0115243&mode=thread&tid=134

Substituting $D = 45$ and $T = 196$ in Eq.(I.1), we obtain that $R \approx 0.23$. The values of *R* for the rest of the dozen studied papers are given in Table I.2.

As we pointed out in Ref. [2] the above reasoning would be convincing if the people who introduced original misprints had always read the original paper. It is more reasonable to assume that the probability of introducing a new misprint in a citation does not depend on whether the author had read the original paper. Then, if the fraction of read citations is *R*, the number of readers in our sample is *RD*, and the ratio of the number of readers to the number of citers in the sample is *RD/T*. What happens to our estimate, Eq. (I.1)? It is correct, just the sample is not representative: the fraction of read citations among the misprinted citations is less than in general citation population.

Can we still determine *R* from our data? Yes. From the misprint statistics we can determine the average number of times, $n_p$, a typical misprint propagates:

$$n_p = \frac{T - D}{D}. \tag{I.2}$$

The number of times a misprint had propagated is the number of times the citation was copied from either the paper which introduced the original misprint, or from one of subsequent papers, which copied (or copied from copied etc) from it. A misprinted citation should not differ from a correct citation as far as copying is concerned. This means that a selected at random citation, on average, is copied (including copied from copied etc) $n_p$ times. The read citations are no different from unread citations as far as copying goes. Therefore, every read citation, on average, was copied $n_p$ times. The fraction of read citations is thus

$$R = \frac{1}{1 + n_p}. \tag{I.3}$$

After substituting Eq. (I.2) into Eq. (I.3), we recover Eq. (I.1).

Note, however, that the average number of times a misprint propagates is not equal to the number of times the citation was copied, but to the number of times it was copied *correctly*. Let us denote the average number of citations copied (including copied from copied etc) from a particular citation as $n_c$. It can be de-termined from $n_p$ the following way. The $n_c$ consists of two parts: $n_p$ (the correctly copied citations) and misprinted citations. If the probability of making a misprint is $M$ and the number of correctly copied citations is $n_p$ then the total number of copied citations is $\dfrac{n_p}{1-M}$ and the number of misprinted citations is $\dfrac{n_p M}{1-M}$. As each misprinted citation was itself copied $n_c$ times, we have the following self-consistency equation for $n_c$:

$$n_c = n_p + n_p \times \frac{M}{1-M} \times (1+n_c) \qquad (I.4)$$

It has the solution

$$n_c = \frac{n_p}{1-M-n_p \times M} \qquad (I.5)$$

After substituting Eq. (I.2) into Eq. (I.5) we get:

$$n_c = \frac{T-D}{D-MT}. \qquad (I.6)$$

From this, we get:

$$R = \frac{1}{1+n_c} = \frac{D}{T} \times \frac{1-(MT)/D}{1-M} \qquad (I.7)$$

The probability of making a misprint we can estimate as $M = \dfrac{D}{N}$, where $N$ is the total number of citations. After substituting this into Eq. (I.7) we get:

$$R = \frac{D}{T} \times \frac{N-T}{N-D}. \qquad (I.8)$$

Substituting $D=45$, $T=196$, and $N=4301$ in Equation (I.8), we get $R \approx 0.22$, which is very close to the initial estimate, obtained using Eq.(I.1).

The values of *R* for the rest of the papers are given in Table I.2. They range between 11% and 58%.

**Table I.2.** Citation and misprint statistics together with estimates of *R* for twelve studied papers. The citation data were retrieved from the ISI database in late 2002 and early 2003. The way we count misprints is look at the whole sequence of volume, page number and the year, which amounts to between 8 and 11 digits for different studied papers. That is, two misprints are distinct if they are different in any of the places, and they are repeats if they agree on all of the digits.

| No. | Citations | Misprints | | M (%) | R | | | | % rank for R = 0.2 |
|---|---|---|---|---|---|---|---|---|---|
| | | total | distinct | | Eq. (I.1) | Eq. (I.8) | Eq. (II.14) | MC | |
| 1 | 1291 | 61 | 29 | 2.2 | 0.48 | 0.46 | 0.44 | 0.37 | 15% |
| 2 | 861 | 33 | 13 | 1.5 | 0.39 | 0.38 | 0.35 | 0.28 | 44% |
| 3 | 818 | 38 | 11 | 1.3 | 0.29 | 0.28 | 0.22 | 0.22 | 68% |
| 4 | 2578 | 263 | 32 | 1.2 | 0.12 | 0.11 | - | 0.10 | 95% |
| 5 | 4301 | 196 | 45 | 1.0 | 0.23 | 0.22 | 0.17 | 0.15 | 76% |
| 6 | 1673 | 40 | 12 | 0.7 | 0.30 | 0.29 | 0.25 | 0.22 | 65% |
| 7 | 1639 | 36 | 21 | 1.3 | 0.58 | 0.58 | 0.57 | 0.49 | 6% |
| 8 | 837 | 55 | 18 | 2.2 | 0.33 | 0.31 | 0.26 | 0.22 | 57% |
| 9 | 419 | 20 | 8 | 1.9 | 0.40 | 0.39 | 0.34 | 0.29 | 50% |
| 10 | 1717 | 33 | 14 | 0.8 | 0.42 | 0.42 | 0.40 | 0.31 | 36% |
| 11 | 1348 | 78 | 27 | 2.0 | 0.35 | 0.33 | 0.29 | 0.23 | 47% |
| 12 | 397 | 61 | 18 | 4.5 | 0.30 | 0.26 | 0.17 | 0.19 | 69% |

In the next section we introduce and solve the stochastic model of misprint propagation. The model explains the power law of misprint repetitions (see Fig.II.1). If you do not have time to read the whole chapter, you can proceed after Section II.1 right to Section III.1. There we formulate and solve the model of random citing scientists (RCS). The model is as follows: when scientist is writing a manuscript he picks up several random papers cites them and copies a fraction of their references. The model can explain why some papers are far more cited than others. After that, you can directly proceed to discussion in Section V. If you have questions, you can find answers to some of them in other sections. The results of Section I.1 are exact in the limit of infinite number of citations. Since this number is obviously finite, we need to study finite size effects, which affect our estimate of *R*. This is done in Section II.2 using complicated mathematical methods and in Section II.3 using Monte Carlo simulations. The limitations of the simple model arising from the instances like, for example, the same author repeats the same misprint, are discussed in Section II.4. In Section II.5, we review the previous work on identical misprints. In short: some people did notice repeat misprints and at-tributed them to citation copying, but nobody derived Eq. (I.1) before us. The RCS model of Section III can explain a power law in overall citation distribution, but cannot explain a power law distribution in citations to the papers of the same age. Section IV.1 introduces the modified model of random-citing scientist (MMRCS), which solves the problem. The model is as follows: when a scientist writes a manuscript, he picks up several random *recent* papers, cites them, and also copies some of their references. The difference with the original model is the word *recent*. In Section IV.2 the MMRCS is solved using theory of branching processes and the power law distribution of citations to the papers of the same age is derived. Section IV.3 considers the model where papers are not created equal but have Darwinian fitness that affects their citability. Section IV.4 studies effects of literature growth (yearly increase of the number of published papers) on citation distribution. Section IV.5 describes numerical simulations of MMRCS, which perfectly match real citation data. Section IV.6 shows that MMRCS can explain the phenomenon of literature aging that is why papers become less cited as they get older. Section IV.7 shows that MMRCS can explain the mysterious phenomenon of sleeping beauties in science (papers that are at first hardly noticed suddenly awake and get a lot of citations). Section IV.8 describes the connection of MMRCS to the Science of Self-Organized Criticality.

## II. Stochastic modeling of misprints in citations

### *1. Misprint propagation model*

Our misprint propagation model (MPM) [1], [3] which was stimulated by Simon's [4] explanation of Zipf Law and Krapivsky-Redner [5] idea of link redirection, is as follows. Each new citer finds the reference to the original in any of the papers that already cite it (or it can be the original paper itself). With probability *R* he gets the citation information from the original. With probability 1-*R* he copies the citation to the original from the paper he found the citation in. In either case, the citer introduces a new misprint with probability *M*.

Let us derive the evolution equations for the misprint distribution. The only way to increase the number of misprints that appeared only once, $N_1$, is to introduce a new misprint. So, with each new citation $N_1$ increases by 1 with probability *M*. The only way to decrease $N_1$, is to copy correctly one of misprints that appeared only once, this happens with probability $\alpha \times \frac{N_1}{N}$, where

$$\alpha = (1-R) \times (1-M) \tag{II.1}$$

is the probability that a new citer copies the citation without introducing a new error, and *N* is the total number of citations. For the expectation value, we thus have:

$$\frac{dN_1}{dN} = M - \alpha \times \frac{N_1}{N}. \tag{II.2}$$

The number of misprints that appeared $K$ times, $N_K$, (where $K > 1$) can be increased only by copying correctly a misprint which appeared $K-1$ times. It can only be decreased by copying (again correctly) a misprint which appeared $K$ times. For the expectation values, we thus have:

$$\frac{dN_K}{dN} = \alpha \times \frac{(K-1) \times N_{K-1} - K \times N_K}{N} \quad (K > 1). \tag{II.3}$$

Assuming that the distribution of misprints has reached its stationary state, we can replace the derivatives ($dN_K/dN$) by ratios ($N_K/N$) to get:

$$\frac{N_1}{N} = \frac{M}{1+\alpha}; \qquad \frac{N_{K+1}}{N_K} = \frac{K}{1+1/\alpha + K} \quad (K > 1). \tag{II.4}$$

Note that for large $K$: $N_{K+1} \approx N_K + dN_K/dK$, therefore Equation (II.4) can be rewritten as:

$$\frac{dN_K}{dK} \approx -\frac{1+1/\alpha}{1+1/\alpha + K} N_k \approx \frac{1+1/\alpha}{K} N_k.$$

From this follows that the misprints frequencies are distributed according to a power law:

$$N_K \sim 1/K^\gamma, \tag{II.5}$$

where

$$\gamma = 1 + \frac{1}{\alpha} = 1 + \frac{1}{(1-R) \times (1-M)}. \tag{II.6}$$

Relationship between $\gamma$ and $\alpha$ in Eq.(II.6) is the same as the one between exponents of number-frequency and rank-frequency distributions[2]. Therefore the parameter $\alpha$, which was defined in Eq.(II.1), turned out to be the Zipf law exponent. An exact formula for $N_k$ can also be obtained by iteration of Eq.(II.4) to get:

$$\frac{N_K}{N} = \frac{\Gamma(K)\Gamma(\gamma)}{\Gamma(K+\gamma)} \times \frac{M}{\alpha} = B(K,\gamma) \times \frac{M}{\alpha} \qquad (II.7)$$

Here $\Gamma$ and $B$ are Euler's Gamma and Beta functions. Using the asymptotic for constant $\gamma$ and large $K$

$$\frac{\Gamma(\gamma)}{\Gamma(K+\gamma)} \sim K^{-\gamma} \qquad (II.8)$$

we recover Eq.(II.5).

The rate equation for the total number of misprints is:

$$\frac{dT}{dN} = M + \alpha \times \frac{T}{N}. \qquad (II.9)$$

The stationary solution of Eq. (II.9) is:

$$\frac{T}{N} = \frac{M}{1-\alpha} = \frac{M}{R+M-RM}. \qquad (II.10)$$

---

[2]Suppose that the number of occurrences of a misprint ($K$), as a function of the rank ($r$), when the rank is determined by the above frequency of occurrence (so that the most popular misprint has rank 1, second most frequent misprint has rank 2 and so on), follows a Zipf law: $K(r) = \frac{C}{r^\alpha}$. We want to find the number-frequency distribution, i.e. how many misprints appeared $n$ times. The number of misprints that appeared between $K_1$ and $K_2$ times is obviously $r_2-r_1$, where $K_1 = C/r_1^\alpha$ and $K_2 = C/r_2^\alpha$. Therefore, the number of misprints that appeared $K$ times, $N_k$, satisfies $N_K dK = -dr$ and hence, $N_K = -dr/dK \sim K^{-1/\alpha-1}$.

The expectation value for the number of distinct misprints is obviously

$$D = N \times M. \tag{II.11}$$

From Equations (II.10) and (II.11) we obtain:

$$R = \frac{D}{T} \times \frac{N-T}{N-D}, \tag{II.12}$$

which is identical to Eq.(I.8).

One can ask why we did not choose to extract $R$ using Equations (II.1) or (II.6). This is because $\alpha$ and $\gamma$ are not very sensitive to $R$ when it is small (in fact Eq. (II.1) gives negative values of $R$ for some of the fittings in Figure II.1). In contrast, $T$ scales as $1/R$.

We can slightly modify our model and assume that original misprints are only introduced when the reference is derived from the original paper, while those who copy references do not introduce new misprints (e.g. they do cut and paste). In this case one can show that $T = N \times M$ and $D = N \times M \times R$. As a consequence Eq.(I.1) becomes exact (in terms of expectation values, of course).

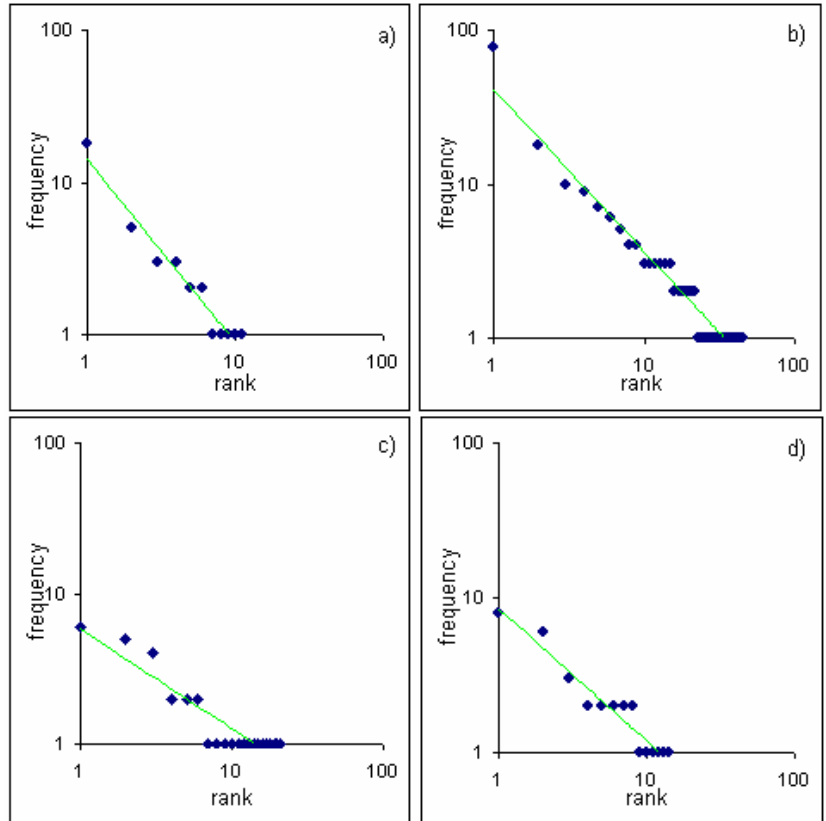

**Figure II.1.** Rank-frequency distributions of misprints in referencing four high-profile papers (here the rank is determined by the frequency so that the most popular misprint has rank 1, second most frequent misprint has rank 2 and so on). Figures a), b), c), and d) are for Papers 2, 5, 7, and 10 of Table I.1. Solid lines are fits to Zipf Law with exponents a) 1.20; b) 1.05; c) 0.66; d) 0.85.

## *2. Finite-size corrections*

Preceding analysis assumes that the misprint distribution had reached its stationary state. Is this reasonable? Eq.(II.9) can be rewritten as:

$$\frac{d(T/N)}{M - (T/N) \times (1-\alpha)} = d \ln N .\tag{II.13}$$

Naturally the first citation is correct (it is the paper itself). Then the initial condition is $N = 1; T = 0$. Eq.(II.13) can be solved to get:

$$\frac{T}{N} = \frac{M}{1-\alpha} \times \left(1 - \frac{1}{N^{1-\alpha}}\right) = \frac{M}{R + M - M \times R} \times \left(1 - \frac{1}{N^{R+M-M \times R}}\right) \tag{II.14}$$

This should be solved numerically for $R$. The values obtained using Eq.(II.14) are given in Table I.2. They range between 17% and 57%. Note that for one paper (No.4) no solution to Eq. (II.14) was found[3]. As $N$ is not a continuous variable, integration of Equation (II.9) is not perfectly justified, particularly when $N$ is small. Therefore, we reexamine the problem using a rigorous discrete approach due to Krapivsky and Redner [6]. The total number of misprints, $T$, is a random variable that changes according to

$$T(N+1) = \begin{cases} T(N) & \text{with probability} \quad 1 - M - \frac{T(N)}{N}\alpha \\ T(N)+1 & \text{with probability} \quad M + \frac{T(N)}{N}\alpha \end{cases} \tag{II.15}$$

after each new citation. Therefore, the expectation values of $T$ obey the following recursion relations:

$$\langle T(N+1) \rangle = \langle T(N) \rangle + \frac{\langle T(N) \rangle}{N}\alpha + M \tag{II.16}$$

To solve Eq. (II.16) we define a generating function:

---

[3] Why did this happen? Obviously, $T$ reaches maximum when $R$ equals zero. Substituting $R = 0$ in Eq.(II.14) we get: $T_{MAX} = N \times \left(1 - 1/N^M\right)$. For paper No.4 we have $N = 2,578$, $M = D/N = 32/2,578$. Substituting this into the preceding equation, we get $T_{MAX} = 239$. The observed value $T = 263$ is therefore higher than an expectation value of $T$ for any $R$. This does not immediately suggest discrepancy between the model and experiment but a strong fluctuation. In fact out of 1,000,000 runs of Monte-Carlo simulation of MPM with the parameters of the mentioned paper and $R$=0.2 exactly 49,712 runs (almost 5%) produced $T \geq 263$.

$$\chi(\omega) = \sum_{N=1}^{\infty} \langle T(N) \rangle \omega^{N-1} \tag{II.17}$$

After multiplying Eq. (II.16) by $N\omega^{N-1}$ and summing over $N \geq 1$ the recursion relation is converted into the differential equation for the generating function

$$(1-\omega)\frac{d\chi}{d\omega} = (1+\alpha)\chi + \frac{M}{(1-\omega)^2} \tag{II.18}$$

Solving Eq.(II.18) subject to the initial condition $\chi(0) = \langle T(1) \rangle = 0$ gives

$$\chi(\omega) = \frac{M}{1-\alpha}\left(\frac{1}{(1-\omega)^2} - \frac{1}{(1-\omega)^{1+\alpha}}\right) \tag{II.19}$$

Finally we expand the right hand side of Eq. (II.19) in Taylor series in $\omega$ and equating coefficients of $\omega^{N-1}$ obtain:

$$\frac{\langle T(N) \rangle}{N} = \frac{M}{1-\alpha}\left(1 - \frac{\Gamma(N+\alpha)}{\Gamma(1+\alpha)\Gamma(N+1)}\right) \tag{II.20}$$

Using Eq. (II.8) we obtain that for large $N$

$$\frac{\langle T(N) \rangle}{N} = \frac{M}{1-\alpha}\left(1 - \frac{1}{\Gamma(1+\alpha)} \times \frac{1}{N^{1-\alpha}}\right) \tag{II.21}$$

This is identical to Eq. (II.14) except for the pre-factor $1/\Gamma(1+\alpha)$. Parameter $\alpha$ (it is defined in Eq. (II.1)) ranges between 0 and 1. Therefore, the argument of Gamma-function ranges between 1 and 2. Because $\Gamma(1) = \Gamma(2) = 1$ and between 1 and 2 Gamma function has just one extremum $\Gamma(1.4616...) = 0.8856...$, the continuum approximation (Eq.(II.14)) is reasonably accurate.

## *3. Monte-Carlo simulations*

In the preceding section, we calculated the expectation value of *T*. However, it does not always coincide with the most likely value when the probability distribution is not Gaussian. To get a better idea of the model's behavior for small *N* and a better estimate of *R* we did numerical simulations. To simplify comparison with actual data the simulations were performed in a "micro-canonical ensemble," i.e. with a fixed number of distinct misprints. Each paper is characterized by the total number of citations, *N*, and the number of distinct misprints, *D*. At the beginning of a simulation, *D* misprints are randomly distributed between *N* citations and chronological numbers of the citations with misprints are recorded in a list. In the next stage of the simulation for each new citation, instead of introducing a misprint with probability *M*, we introduce a misprint only if its chronological number is included in the list created at the outset. This way one can ensure that the number of distinct misprints in every run of a simulation is equal to the actual number of distinct misprints for the paper in question. A typical outcome of such simulation for Paper No.5 is shown in Fig. II.2.

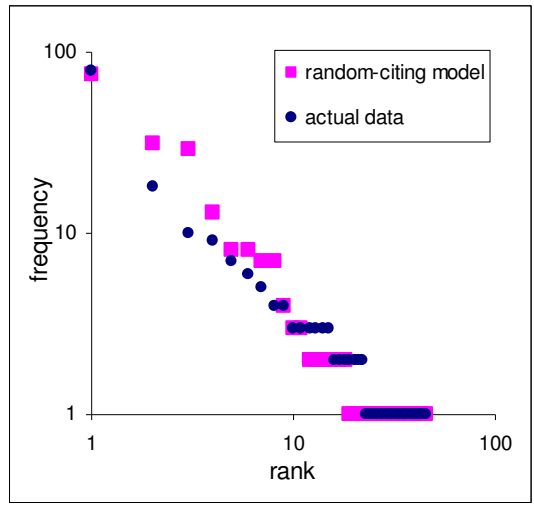

**Figure II.2.** A typical outcome of a single simulation of the MPM (with *R*=0.2) compared to the actual data for Paper No.5 in Table I.1.

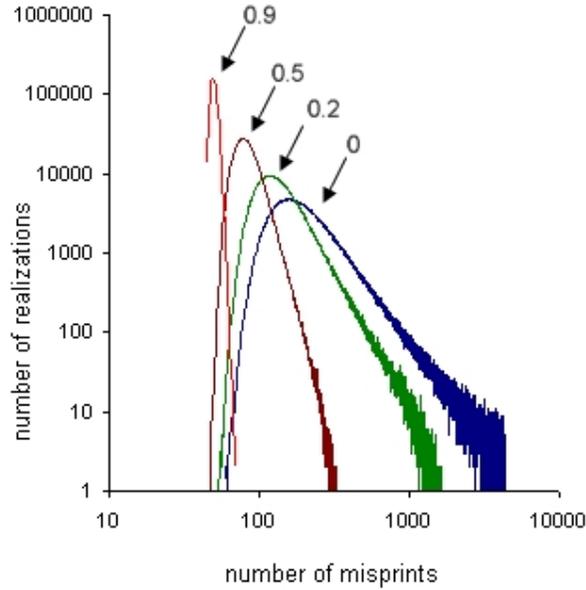

**Figure II.3.** The outcome of 1,000,000 runs of the MPM with $N$=4301, $D$=45 (parameters of paper No.5 from Table I.1) for four different values of $R$ (0.9, 0.5, 0.2, 0 from left to right).

To estimate the value of $R$, 1,000,000 runs of the random-citing model with $R$ = 0, 0.1, 0.2… 0.9 were done. An outcome of such simulations for one paper is shown in Fig. II.3. The number of times, $N_R$, when the simulation produced a total number of misprints equal to the one actually observed for the paper in question was recorded for each $R$. Bayesian inference was used to estimate the probability of $R$:

$$P(R) = \frac{N_R}{\sum_R N_R} \tag{II.22}$$

Estimated probability distributions of $R$, computed using Eq. (II.22) for four sample papers are shown in Figure II.4. The median values are given in Table I.2 (see the MC column). They range between 10% and 49%.

Now let us assume $R$ to be the same for all twelve papers and compute Bayesian inference:

$$P(R) = \frac{\prod_{i=1}^{12} N_R^i}{\sum_R \prod_{i=1}^{12} N_R^i} \tag{II.23}$$

The result is shown in Fig. II.5. $P(R)$ is sharply peaked around $R=0.2$. The median value of $R$ is 18% and with 95% probability $R$ is less than 34%.

But is the assumption that $R$ is the same for all twelve papers reasonable? The estimates for separate papers vary between ten and fifty percent! To answer this question we did the following analysis. Let us define for each paper a "percentile rank." This is the fraction of the simulations of the MPM (with $R = 0.2$) that produced $T$, which was less than actually observed $T$. Actual values of these percentile ranks for each paper are given in Table I.2 and their cumulative distribution is shown in Fig. II.6. Now if we claim that MPM with same $R = 0.2$ for all papers indeed describes the reality – then the distribution of these percentile ranks must be uniform. Whether or not the data is consistent with this, we can check using Kolmogorov-Smirnov test [7]. The maximum value of the absolute difference between two cumulative distribution functions ($D$-statistics) in our case is $D = 0.15$. The probability for $D$ to be more than that is 91%. This means that the data is perfectly consistent with the assumption of $R = 0.2$ for all papers.

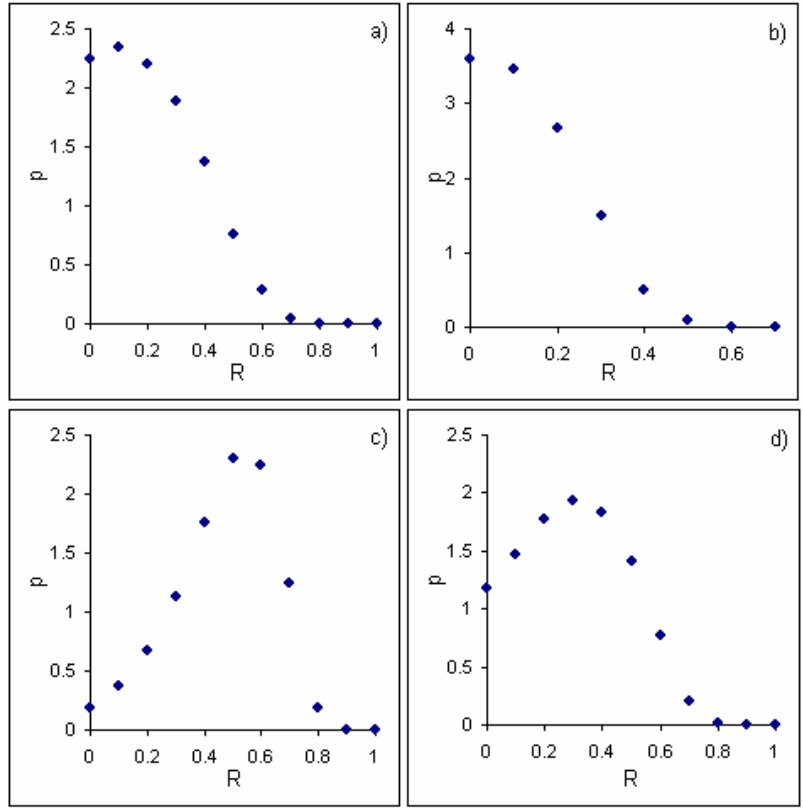

**Figure II.4.** Bayesian inference for the probability density of the readers/citers ratio, R, computed using Eq. (II.22). Figures a), b), c), and d) are for Papers No. 2, 5, 7, and 10 (Table I.1).

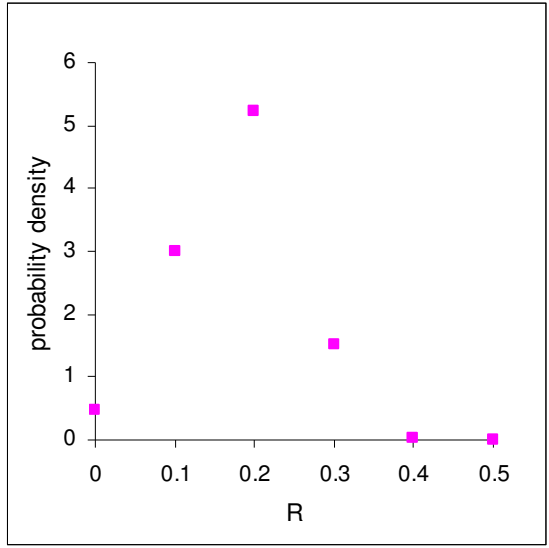

**Figure II.5.** Bayesian inference for the readers/citers ratio, *R*, based on twelve studied papers computed using Eq.(II.23).

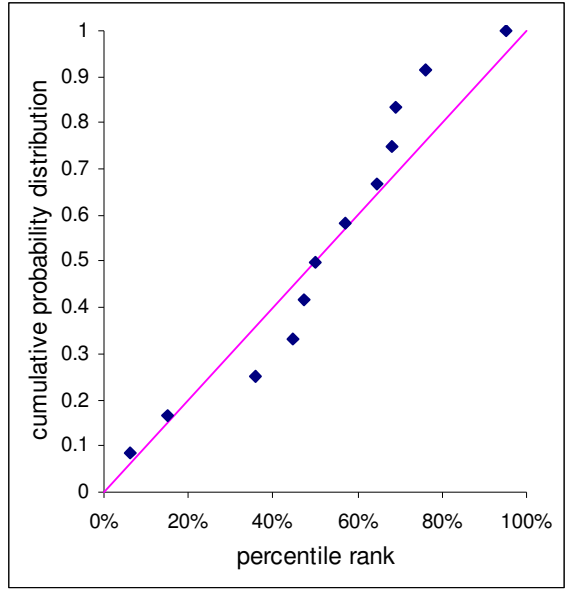

**Figure II.6.** Cumulative distribution of the percentile ranks of the observed values of $T$ with regard to the outcomes of the simulations of the MPM with $R = 0.2$ (diamonds). For comparison the cumulative function of the uniform distribution is given (a line).

One can notice that the estimates of $M$ (computed as $M=D/N$) for different papers (see Table I.2) are also different. One may ask if it is possible that $M$ is the same for all papers and different values of $D/N$ are results of fluctuations. The answer is that the data is totally inconsistent with single $M$ for all papers. This is not unexpected, because some references can be more error-prone, for example, because they are longer. Indeed, the most-misprinted paper (No.12) has two-digit volume number and five-digit page number.

## *4. Operational limitations of the model*

Scientists copy citations because they are not perfect. Our analysis is imperfect as well. There are occasional repeat identical misprints in papers, which share individuals in their author lists. To estimate the magnitude of this effect we took a close look at all 196 misprinted citations to paper 5 of Table I.1. It turned out that such events constitute a minority of repeat misprints. It is not obvious what to do with such cases when the author lists are not identical: should the set of citations be counted as a single occurrence (under the premise that the common co-author is the only source of the misprint) or as multiple repetitions. Counting all such repetitions as only a single misprint occurrence results in elimination of 39 repeat misprints. The number of total misprints, $T$, drops from 196 to 157, bringing the upper bound for $R$ (Eq.(I.1)) from $\frac{45}{196} \cong 23\%$ up to $\frac{45}{157} \cong 29\%$. An alternative approach is to subtract all the repetitions of each misprint by the originators of that misprint from non-readers and add it to the number of readers. There were 11 such repetitions, which increases $D$ from 45 up to 56 and the upper bound for $R$ (Eq.(I.1)) rises to $\frac{56}{196} \cong 29\%$, which is the same value as the preceding estimate. It would be desirable to redo the estimate using Equations (II.12) and (II.14), but the misprint propagation model would have to be modified to account for repeat citations by same author and multiple authorships of a paper. This may be a subject of future investigations.

Another issue brought up by the critics [8] is that because some misprints are more likely than others, it is possible to repeat someone else's misprint purely by chance. By examining the actual data, one finds that about two third of distinct misprints fall in to the following categories:
  a) One misprinted digit in volume, page number, or in the year.
  b) One missing or added digit in volume or page number.
  c) Two adjacent digits in a page number are interchanged.

The majority of the remaining misprints are combinations of a), b), c), for example, one digit in page number omitted and one digit in year misprinted[4]. For a typical reference, there are over fifty aforementioned likely misprints. However, even if probability of certain misprint is not negligibly small but one in fifty, our analysis still applies. For example, for paper No.5 (Table I.1) the most popular error appeared 78 times, while there were 196 misprints in total. Therefore, if probability of certain misprint is 1/50, there should be about $196/50 \approx 4$ such misprints, not 78. In order to explain repeat misprints distribution by higher probability of certain misprint this probability should be as big as $78/196 \approx 0.4$. This is extremely unlikely. However, finding relative propensities of different misprints deserves further investigation.

Smith noticed [9] that some misprints are in fact introduced by the ISI. To estimate the importance of this effect we explicitly verified 88 misprinted (according to ISI) citations in the original articles. 72 of them were exactly as in the ISI database, but 16 were in fact correct citations. To be precise some of them had minor inaccuracies, like second initial of the author was missing, while page number, volume and year where correct. Apparently, they were victims of an "erroneous error correction" [9]. It is not clear how to consistently take into account these effects, specifically because there is no way to estimate how many wrong citations have been correctly corrected by ISI [10]. But given the relatively small percentage of the discrepancy between ISI database and actual articles ($16/88 \cong 18\%$) this can be taken as a noise with which we can live.

---

[4]There are also misprints where author, journal, volume and year are perfectly correct, but the page number is totally different. Probably, in such case the citer mistakenly took the page number from a neighboring paper in the reference list he was lifting the citation from.

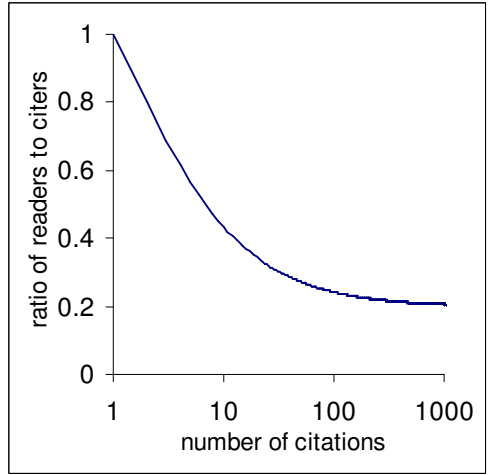

**Figure II.7.** Ratio of readers to citers as a function of total amount of citations for $R = 0.2$, computed using Eq.(II.24).

It is important to note that within the framework of the MPM $R$ is not the ratio of readers to citers, but the probability that a citer consults the original paper, provided that he encountered it through another paper's reference list. However, he could encounter the paper directly. This has negligible effect for highly-cited papers, but is important for low-cited papers. Within the MPM framework the probability of such an event for each new citation is obviously $1/n$, where $n$ is the current total number of citations. The expectation value of the true ratio of readers to citers is therefore:

$$R^*(N) = R + (1-R) \times \frac{\sum_{n=1}^{N} \frac{1}{n}}{N} \approx R + (1-R) \times \frac{\ln(2N+1)}{N}. \tag{II.24}$$

The values of $R^*$ for papers with different total numbers of citations, computed using Eq.(II.24), are shown in Fig.II.7. For example, on average, about four people have read a paper which was cited ten times. One can use Eq.(II.24) and empirical citation distribution to estimate an average value of $R^*$ for the scientific literature in general. The formula is:

$$\left\langle R^* \right\rangle = \frac{\sum R^*(N_i) \times N_i}{\sum N_i} \tag{II.25}$$

Here the summation is over all of the papers in the sample and $N_i$ is the number of citations that *i*th paper had received. The estimate, computed using citation data for Physical Review D [11] and Eqs (II.24) and (II.25) (assuming $R = 0.2$), is $\left\langle R^* \right\rangle \approx 0.33$.

## *5. Comparison with the previous work*

The bulk of previous literature on citations was concerned with their counting. After extensive literature search we found only a handful of papers which analyzed misprints in citations (the paper by Steel [12], titled identically to our first misprint paper, i.e. "Read before you cite," turned out to use the analysis of the content of the papers, not of the propagation of misprints in references). Broadus [13] looked through 148 papers, which cited both the renowned book, which misquoted the title of one of its references, and that paper, the title of which was misquoted in the book. He found that 34 or 23% of citing papers made the same error as was in the book. Moed and Vries [14] (apparently independent of Broadus, as they don't refer to his work), found identical misprints in scientific citations and attributed them to citation copying. Hoerman and Nowicke [15] looked through a number of papers, which deal with the so-called Ortega Hypothesis of Cole and Cole. When Cole and Cole quoted a passage from the book by Ortega they introduced three distortions. Hoerman and Nowicke found seven papers which cite Cole and Cole and also quote that passage from Ortega. In six out of these seven papers all of the distortions made by Cole and Cole were repeated. According to [15] in this process even the original meaning of the quotation was altered. In fact, information is sometimes *defined* by its property to deteriorate in chains [16].

While the fraction of copied citations found by Hoerman and Nowicke [15], $6/7 \cong 86\%$ agrees with our estimate, Boadus' number, 23%, seems to disagree with it. Note, however, that Broadus [13] assumes that citation, if copied - was copied from the book (because the book was renowned). Our analysis indicates that majority of citations to renowned papers are copied. Similarly, we sur-mise, in the Broadus' case citations to both the book and the paper were often copied from a third source.

### III. Copied citations create renowned papers?

*1. The model of random-citing scientists*

During the "Manhattan project" (the making of nuclear bomb), Fermi asked Gen. Groves, the head of the project, what is the definition of a "great" general [17]. Groves replied that any general who had won five battles in a row might safely be called great. Fermi then asked how many generals are great. Groves said about three out of every hundred. Fermi conjectured that considering that opposing forces for most battles are roughly equal in strength, the chance of winning one battle is ½ and the chance of winning five battles in a row is $1/2^5 = 1/32$. "So you are right, General, about three out of every hundred. Mathematical probability, not genius." The existence of military genius also questioned Lev Tolstoy in his book "War and Peace."

A commonly accepted measure of "greatness" for scientists is the number of citations to their papers [18]. For example, SPIRES, the High-Energy Physics literature database, divides papers into six categories according to the number of citations they receive. The top category, "Renowned papers" are those with 500 or more citations. Let us have a look at the citations to roughly eighteen and a half thousand papers[5], published in Physical Review D in 1975-1994 [11]. As of 1997 there where about 330 thousands of such citations: eighteen per published paper on average. However, forty-four papers were cited five hundred times or more. Could this happen if **all papers are created equal**? If they indeed are then the chance to win a citation is one in 18,500. What is the chance to win 500 cites out of 330,000? The calculation is slightly more complex than in the militaristic case[6], but the answer is 1 in $10^{500}$, or, in other words, it is zero. One is tempted to conclude that those forty-four papers, which achieved the impossible, are great.

---

[5] In our initial report [22] we mentioned "over 24 thousand papers". This number is incorrect and the reader surely understands the reason: misprints. In fact, out of 24,295 "papers" in that dataset only 18,560 turned out to be real papers and 5,735 "papers" turned out to be misprinted citations. These "papers" got 17,382 out of 351,868 citations. That is every distinct misprint on average appeared three times. As one could expect, cleaning out misprints lead to much better agreement between experiment and theory: compare Fig. III.1 and Fig. 1 of Ref [22].

In the preceding sections, we demonstrated that copying from the lists of references used in other papers is a major component of the citation dynamics in scientific publication. This way a paper that already was cited is likely to be cited again, and after it is cited again it is even more likely to be cited in the future. In other words, "unto every one which hath shall be given" [Luke 19:26]. This phenomenon is known as "Matthew effect,"[7] "cumulative advantage" [20], or "preferential attachment" [21].

The effect of citation copying on the probability distribution of citations can be quantitatively understood within the framework of **the model of random-citing scientists (RCS)** [22][8], which is as follows. When a scientist is writing a manuscript he picks up $m$ random articles[9], cites them, and also copies some of their references, each with probability $p$.

---

[6] If one assumes that all papers are created equal then the probability to win $m$ out of $n$ possible citations when the total number of cited papers is $N$ is given by the Poisson distribution: $p = \frac{(n/N)^m}{m!} \times e^{-n/N}$. Using Stirling's formula one can rewrite this as: $\ln(p) \approx m \ln\left(\frac{ne}{Nm}\right) - \frac{n}{N}$. After substituting $n = 330,000$, $m = 500$ and $N = 18500$ into the above equation we get: $\ln(p) \approx -1,180$, or $p \approx 10^{-512}$.

[7] Sociologist of science Robert Merton observed [19] that when a scientist gets recognition early in his career he is likely to get more and more recognition. He called it "Matthew Effect" because in Gospel according to Mathew (25:29) appear the words: "unto every one that hath shall be given." The attribution of a special role to St. Matthew is unfair. The quoted words belong to Jesus and also appear in Luke and Mark's gospels. Nevertheless, thousands of people who did not read The Bible copied the name "Matthew Effect."

[8] From the mathematical perspective, almost identical to RCS model (the only difference was that they considered an undirected graph, while citation graph is directed) was earlier proposed in Ref. [27].

[9]The analysis presented here also applies to a more general case when $m$ is not a constant, but a random variable. In that case $m$ in all of the equations that follow should be interpreted as the mean value of this variable.

The evolution of the citation distribution (here $N_K$ denotes the number of papers that were cited $K$ times, and $N$ is the total number of papers) is described by the following rate equations:

$$\frac{dN_0}{dN} = 1 - m \times \frac{N_0}{N}, \tag{III.1}$$

$$\frac{dN_K}{dN} = m \times \frac{(1+p(K-1))N_{K-1} - (1+pK)N_K}{N},$$

which have the following stationary solution:

$$N_0 = \frac{N}{m+1}; \quad N_K = \frac{1+p(K-1)}{1+1/m+pK} N_{K-1}. \tag{III.2}$$

For large $K$ it follows from (III.2) that:

$$N_K \sim 1/K^\gamma; \quad \gamma = 1 + \frac{1}{m \times p}. \tag{III.3}$$

Citation distribution follows a power law, empirically observed in [23], [24], [25].

A good agreement between the RCS model and actual citation data [11] is achieved with input parameters $m = 5$ and $p = 0.14$ (see Figure III.1). Now what is the probability for an arbitrary paper to become "renowned," i.e. receive more than five hundred citations? Iteration of Eq. III.2 (with $m = 5$ and $p = 0.14$) shows that this probability is 1 in 420. This means that about 44 out of 18,500 papers should be renowned. Mathematical probability, not genius.

On one incident [26] Napoleon (incidentally, he was the military commander, whose genius was questioned by Tolstoy) said to Laplace "They tell me you have written this large book on the system of the universe, and have never even mentioned its Creator." The reply was "I have no need for this hypothesis." It is worthwhile to note that Laplace was not against God. He simply did not need to postulate His existence in order to explain existing astronomical data. Similarly, the present work is not blasphemy. Of course, in some spiritual sense, great scientists do exist. It is just that even if they would not exist, citation data would look the same.

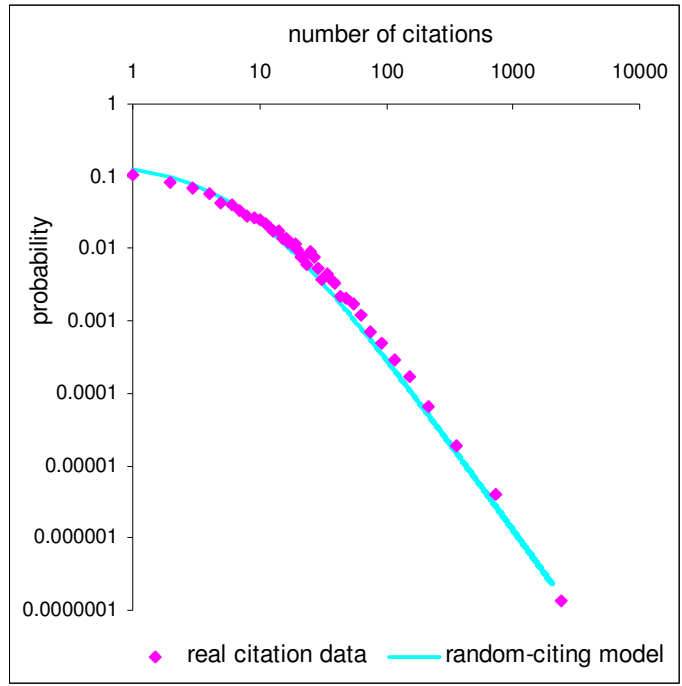

**Figure III.1.** Outcome of the model of random-citing scientists (with $m = 5$ and $p = 0.14$) compared to actual citation data. Mathematical probability rather than genius can explain why some papers are cited a lot more than the others.

## 2. Relation to previous work

Our original paper on the subject [22] was stimulated by the model introduced by Vazquez [26]. It is as follows. When scientist is writing a manuscript, he picks up a paper, cites it, follows its references, and cites a fraction $p$ of them. Afterward he repeats this procedure with each of the papers that he cited. And so on. In two limiting cases ($p = 1$ and $p = 0$) the Vazquez model is exactly solvable [26]. Also in these cases it is identical to the RCS model ($m = 1$ case), which in contrast can be solved for any $p$. Though theoretically interesting, the Vazquez model cannot be a realistic description of the citation process. In fact, the results presented in two preceding sections indicate that there is essentially just one "recursion," that is, references are copied from the paper at hand, but hardly followed. To be precise, results of two preceding sections could support a generalized Vazquez model, in which the references of the paper at hand are copied with probability $p$, and afterwards the copied references are followed with probability *R*. However, given the low value of this probability ($R \approx 0.2$), it is clear that the effect of secondary recursions on the citation distribution is small.

The book of Ecclesiastes says: "Is there any thing whereof it may be said, See, this is new? It hath been already of old time, which was before us." The discovery reported in this section is no exception. Long ago Price [20], by postulating that the probability of paper being cited is somehow proportional to the amount of citations it had already received, explained the power law in citation frequencies, which he had earlier observed [22]. However, Price did not propose any mechanism for that. Vasquez did propose a mechanism, but it was only a hypothesis. In contrast, our paper is rooted in facts.

## IV. Mathematical theory of citing

### 1. Modified model of random-citing scientists

"... citations not only vouch for the authority and relevance of the statements they are called upon to support; they embed the whole work in context of previous achievements and current aspirations. It is very rare to find a reputable paper that contains no reference to other research. Indeed, one relies on the citations to show its place in the whole scientific structure just as one relies on a man's kinship affiliations to show his place in his tribe."

John M. Ziman, FRS (Ref. [28])

In spite of its simplicity, the model of random citing scientists appeared to account for the major properties of empirically observed distributions of citations. A more detailed analysis, however, reveals that some features of the citation distribution are not accounted for by the model. The cumulative advantage process would lead to oldest papers being most highly cited [5], [21], [29][10]. In reality, the average citation rate decreases as the paper in question gets older [23], [30], [31], [32]. The cumulative advantage process would also lead to an exponential distribution of citations to papers of the same age [5], [29]. In reality citations to papers published during the same year are distributed according to a power-law (see the ISI dataset in Fig.1(a) in Ref. [25]).

---

[10] Some of these references do not deal with citing, but with other social processes, which are modeled using the same mathematical tools. Here we rephrase the results of such papers in terms of citations for simplicity.

In this section, we study the ***modified*** model of random-citing scientists [33]: when a scientist writes a manuscript, he picks up several random *recent* papers, cites them and also copies some of their references. The difference with the original model is the word *recent*. We solve this model using methods of the theory of branching processes [34] (we review its relevant elements in Appendix A), and show that it explains both the power-law distribution of citations to papers published during the same year and literature aging. A similar model was earlier proposed by Bentley, Hahn & Shennan [35] in the context of patents citations. However they just used it to explain a power law in citation distribution (for what the usual cumulative advantage model will do) and did not address the topics we just mentioned.

While working on a paper, a scientist reads current issues of scientific journals and selects from them the references to be cited in it. These references are of two sorts:
- *Fresh papers he had just read* – to embed his work in the context of current aspirations.
- *Older papers that are cited in the fresh papers* he had just read – to place his work in the context of previous achievements.

It is not a necessary condition for the validity of our model that the citations to old papers are copied, but the paper itself remains unread (although such opinion is supported by the studies of misprint propagation). The necessary conditions are as follows:
- Older papers are considered for possible citing only if they were recently cited.
- If a citation to an old paper is followed and the paper is formally read – the scientific qualities of that paper do not influence its chance of being cited.

A reasonable estimate for the length of time a scientist works on a particular paper is one year. We will thus assume that "recent" in the model of random-citing scientists means the preceding year. To make the model mathematically tractable we enforce time-discretization with a unit of one year. The precise model to be studied is as follows. Every year *N* papers are published. There is, on average, $N_{ref}$ references in a published paper (the actual value is somewhere between 20 and 40). Each year, a fraction $\alpha$ of references goes to randomly selected preceding year papers (the estimate[11] from actual citation data is $\alpha \approx 0.1$ (see Fig. 4 in Ref. [23]) or $\alpha \approx 0.15$ (see Fig. 6 in Ref. [36]). The remaining citations are randomly copied from the lists of references used in the preceding year papers.

---

[11] The uncertainty in the value of $\alpha$ depends not only on the accuracy of the estimate of the fraction of citations which goes to previous year papers. We also arbitrarily defined recent paper (in the sense of our model), as the one published within a year. Of course, this is by order of magnitude correct but the true value can be anywhere between half a year and two years.

## *2. Branching citations*

When *N* is large, this model leads to the first-year citations being Poisson-distributed. The probability to get *n* citations is

$$p(n) = \frac{\lambda_0^n}{n!} e^{-\lambda_0},  \quad\quad\quad (IV.1)$$

where $\lambda_0$ is the average expected number of citations

$$\lambda_0 = \alpha N_{ref}. \quad\quad\quad (IV.2)$$

The number of the second-year citations, generated by each first year citation (as well as, third-year citations generated by each second year citation and so on), again follows a Poisson distribution, this time with the mean

$$\lambda = (1-\alpha). \quad\quad\quad (IV.3)$$

Within this model, citation process is a branching process (see Appendix A) with the first year citations equivalent to children, the second-year citations to grand children, and so on.

As $\lambda < 1$, this branching process is subcritical. Figure IV.1 shows a graphical illustration of the branching citation process.

Substituting Eq.(IV.1) into Eq.(A1) we obtain the generating function for the first year citations:

$$f_0(z) = e^{(z-1)\lambda_0}. \quad\quad\quad (IV.4)$$

Similarly, the generating function for the later-years citations is:

$$f(z) = e^{(z-1)\lambda}. \quad\quad\quad (IV.5)$$

The process is easier to analyze when $\lambda = \lambda_0$, or $\frac{\lambda_0}{\lambda} = \frac{\alpha}{1-\alpha} N_{ref} = 1$, as then we have a simple branching process, where all generations are governed by the same offspring probabilities. The case when $\lambda \neq \lambda_0$ we study in Appendix B.

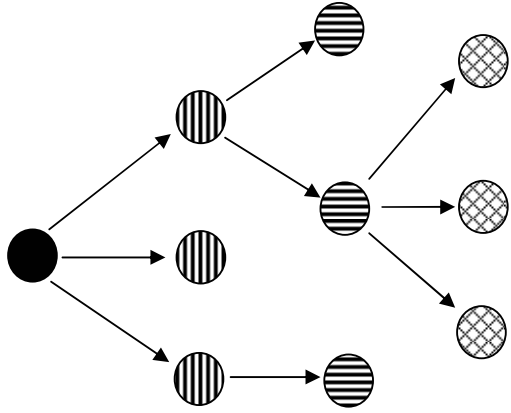

**Figure IV.1.** An illustration of the branching citation process, generated by the modified model of random-citing scientists. During the first year after publication, the paper was cited in three other papers written by the scientists who have read it. During the second year one of those citations was copied in two papers, one in a single paper and one was never copied. This resulted in three second year citations. During the third year, two of these citations were never copied, and one was copied in three papers.

**A. Distribution of citations to papers published during the same year**

Theory of branching processes allows us to analytically compute the probability distribution, $P(n)$, of the total number of citations the paper receives before it is forgotten. This should approximate the distribution of citations to old papers. Substituting Eq.(IV.5) into Eq.(A8) we get:

$$P(n) = \frac{1}{n!}\left[\frac{d^{n-1}}{d\omega^{n-1}}e^{n(\omega-1)\lambda}\right]_{\omega=0} = \frac{(n\lambda)^{n-1}}{n!}e^{-\lambda n} \qquad (IV.6)$$

Applying Stirling's formula to Eq.(IV.6), we obtain the large $n$ asymptotic of the distribution of citations:

$$P(n) \propto \frac{1}{\sqrt{2\pi}\lambda n^{3/2}} e^{-(\lambda-1-\ln\lambda)n} \qquad \text{(IV.7)}$$

When $1-\lambda \ll 1$ we can approximate the factor in the exponent as:

$$\lambda - 1 - \ln\lambda \approx (1-\lambda)^2/2. \qquad \text{(IV.8)}$$

As $1-\lambda \ll 1$, the above number is small. This means that for $n \ll 2/(1-\lambda)^2$ the exponent in Eq.(IV.7) is approximately equal to 1 and the behavior of $P(n)$ is dominated by the $\frac{1}{n^{3/2}}$ factor. In contrast, when $n \gg 2/(1-\lambda)^2$ the behavior of $P(n)$ is dominated by the exponential factor. Thus citation distribution changes from a power law to an exponential (suffers an exponential cut-off) at about

$$n_c = \frac{1}{\lambda - 1 - \ln\lambda} \approx \frac{2}{(1-\lambda)^2} \qquad \text{(IV.9)}$$

citations. For example, when $\alpha = 0.1$ Eq.(IV.3) gives $\lambda = 0.9$ and from Eq.(IV.9) we get that the exponential cut-off happens at about 200 citations. We see that, unlike the cumulative advantage model, our model is capable of qualitative explanation of the power law distribution of citations to papers of the same age. The exponential cut-off at 200, however, happens too soon, as the actual citation distribution obeys a power law well into thousands of citations. In the following sections we show that taking into account the effects of literature growth and of variations in papers' Darwinian fitness can fix this.

In the cumulative advantage (AKA preferential attachment) model, a power law distribution of citations is only achieved because papers have different ages. This is not immediately obvious from the early treatments of the problem [4], [20], but is explicit in later studies [5], [21], [29]. In that model, the oldest papers are the most cited ones. The number of citations is mainly determined by paper's age. At the same time, distribution of citations to papers of the same age is exponential [5], [29]. The key difference between that model and ours is as follows. In the cumulative advantage model, the rate of citation is proportional to the number of citations the paper had accumulated since its publication. In our model, the rate of citation is proportional to the number of citations the paper received during preceding year. This means that if an unlucky paper was not cited during previous year – it will never be cited in the future. This means that its rate of citation will be less than that in the cumulative advantage model. On the other hand, the lucky papers, which were cited during the previous year, will get all the citation share of the unlucky papers. Their citation rates will be higher than in the cumulative advantage model. There is thus more stratification in our model than in the cumulative advantage model. Consequently, the resulting citation distribution is far more skewed.

**B. Distribution of citations to papers *cited* during the same year**

We denote as $N(n)$ the number of papers cited *n* times during given year. The equilibrium distribution of $N(n)$ should satisfy the following self-consistency equation:

$$N(n) = \sum_{m=1}^{\infty} N(m) \frac{(\lambda m)^n}{n!} e^{-\lambda m} + N \frac{(\lambda_0)^n}{n!} e^{-\lambda_0} \qquad (IV.10)$$

Here the first term comes from citation copying and the second from citing previous year papers. In the limit of large *n* the second term can be neglected and the sum can be replaced with integral to get:

$$N(n) = \frac{1}{n!} \int_0^\infty dm N(m)(\lambda m)^n e^{-\lambda m} \qquad (IV.11)$$

In the case $\lambda = 1$ one solution of Eq.(IV.11) is $N(m) = C$, where *C* is an arbitrary constant. Clearly, the integral becomes a gamma-function and the factorial in the denominator cancels out. However, this solution is, meaningless since the total number of citations per year, which is given by

$$N_{cite} = \sum_{m=1}^{\infty} m N(m) \qquad (IV.12)$$

diverges. In the case $\lambda < 1$, $N(m) = C$ is no longer a solution since the integral gives $C/\lambda$. However $N(m) = C/m$ is a solution. This solution is again meaningless because the total number of yearly citations given by (IV.12) again diverges. One can look for a solution of the form

$$N(m) = \frac{C}{m} \exp(-\mu m). \qquad (IV.13)$$

After substituting Eq. (IV.13) into (IV.11) we get that $N(n)$ is given by the same function but instead of $\mu$ with

$$\mu' = \ln(1 + \mu/\lambda). \tag{IV.14}$$

The self consistency equation for $\mu$ is thus

$$\mu = \ln(1 + \mu/\lambda). \tag{IV.15}$$

The obvious solution is $\mu = 0$ which gives us the previously rejected solution $N(m) = C/m$. It is also easy to see that this stationary solution is unstable. If $\mu$ slightly deviates from zero Eq. (IV.14) gives us $\mu' = \mu/\lambda$. Since $\lambda < 1$ the deviation from stationary shape will increase the next year. Another solution of Eq. (IV.15) can be found by expansion of the logarithm up to the second order in $\mu$. It is $\mu \approx 2(1-\lambda)$. One can show that this solution is stable. Thus, we get:

$$N(m) \approx \frac{C}{m} \exp(-2(1-\lambda)m). \tag{IV.16}$$

After substituting this into Eq. (IV.12) we get

$$C \approx 2(1-\lambda)N_{cite}. \tag{IV.17}$$

The solution which we just presented was stimulated by that obtained by Wright [37], who studied the distribution of alleles (alternative forms of a gene) in a population. In Wright's model, the gene pool at any generation has constant size $N_g$. To form the next generation we $N_g$ times select a random gene from current generation pool and copy it to next generation pool. With some probability, a gene can mutate during the process of copying. The problem is identical to ours with an allele replaced with a paper and mutation with a new paper. Our solution follows that of Wright but is a lot simpler. Wright considered finite $N_g$. and as a consequence got Binomial distribution and a Beta function in his analog of Eq.(IV.11). The simplification was possible because in the limit of large $N_g$ Binomial distribution becomes Poissonian. Alternative derivations of Eq.(IV.16) can be found in Refs [33] and [38].

## *3. Scientific Darwinism*

Now we proceed to investigate the model, where papers are not created equal, but each has a specific *Darwinian fitness*, which is a bibliometric measure of scientific fangs and claws that help a paper to fight for citations with its competitors. While this parameter can depend on factors other than the intrinsic quality of the paper, the fitness is the only channel through which the quality can enter our model. The fitness may have the following interpretation. When a scientist writes a manuscript he needs to include in it a certain number of references (typically between 20 and 40, depending on implicit rules adopted by a journal where the paper is to be submitted). He considers random scientific papers one by one for citation, and when he has collected the required number of citations, he stops. Every paper has specific probability to be selected for citing, once it was considered. We will call this probability a Darwinian fitness of the paper. Defined in such way, fitness is bounded between 0 and 1.

In this model a paper with fitness $\phi$ will on average have

$$\lambda_0(\varphi) = \alpha N_{ref}\, \varphi / \langle \varphi \rangle_p \tag{IV.18}$$

first-year citations. Here we have normalized the citing rate by the *average fitness of published papers*, $\langle \varphi \rangle_p$, to insure that the fraction of citations going to previous year papers remained $\alpha$. The fitness distribution of references is different from the fitness distribution of published papers, as papers with higher fitness are cited more often. This distribution assumes an asymptotic form $p_r(\varphi)$, which depends on the distribution of the fitness of published papers, $p_p(\varphi)$, and other parameters of the model.

During later years there will be on average

$$\lambda(\varphi) = (1-\alpha)\varphi / \langle \varphi \rangle_r \tag{IV.19}$$

next year citations per one current year citation for a paper with fitness $\phi$. Here, $\langle \varphi \rangle_r$ is the *average fitness of a reference*.

### A. Distribution of citations to papers published during the same year

Let us start with the self-consistency equation for $p_r(\varphi)$, the equilibrium fitness distribution of references:

$$p_r(\varphi) = \alpha \frac{\varphi \times p_p(\varphi)}{\langle\varphi\rangle_p} + (1-\alpha)\frac{\varphi \times p_r(\varphi)}{\langle\varphi\rangle_r} \tag{IV.20}$$

solution of which is:

$$p_r(\varphi) = \frac{\alpha \times \varphi \times p_p(\varphi)/\langle\varphi\rangle_p}{1 - (1-\alpha)\varphi/\langle\varphi\rangle_r} \tag{IV.21}$$

One obvious self-consistency condition is that

$$\int p_r(\varphi)d\varphi = 1. \tag{IV.22}$$

Another is:

$$\int \varphi \times p_r(\varphi)d\varphi = \langle\varphi\rangle_r.$$

However, when the condition of Eq. (IV.22) is satisfied the above equation follows from Eq. (IV.20).

Let us consider the simplest case when the fitness distribution, $p_p(\varphi)$, is uniform between 0 and 1. This choice is arbitrary, but we will see that the resulting distribution of citations is close to the empirically observed one. In this case, the average fitness of a published paper is $\langle\varphi\rangle_p = 0.5$. After substituting this into Eq. (IV.20), the result into Eq. (IV.22), and performing integration we get:

$$\alpha = -\frac{((1-\alpha)/\langle\varphi\rangle_r)^2/2}{\ln(1-(1-\alpha)/\langle\varphi\rangle_r) + (1-\alpha)/\langle\varphi\rangle_r} \tag{IV.23}$$

Since $\alpha$ is close to 0, $\langle\varphi\rangle_r$ must be very close to $1-\alpha$, and we can replace it with the latter everywhere but in the logarithm to get:

$$\frac{1-\alpha}{\langle\varphi\rangle_r} = 1 - e^{-\frac{1}{2\alpha}-1} \tag{IV.24}$$

For papers of fitness $\varphi$, citation distribution is given by Eq. (IV.6) or Eq. (IV.7) with $\lambda$ replaced with $\lambda(\varphi)$, given by Eq.(IV.19):

$$P(n,\varphi) \propto \frac{1}{\sqrt{2\pi \lambda(\varphi)n^{3/2}}} e^{-(\lambda(\varphi)-1-\ln \lambda(\varphi))n} . \qquad (IV.25)$$

When $\alpha = 0.1$, Eq. (IV.24) gives $(1-\alpha)/\langle\varphi\rangle_r \approx 1 - 2 \times 10^{-3}$. From Eq. (IV.19) it follows that $\lambda(1) = (1-\alpha)/\langle\varphi\rangle_r$. Substituting this into Eq. (IV.9) we get that the exponential cutoff for the fittest papers ($\varphi = 1$) starts at about 300,000 citations. In contrast, for the unfit papers the cut-off is even stronger than in the model without fitness. For example, for papers with fitness $\varphi = 0.1$ we get $\lambda(0.1) = 0.1(1-\alpha)/\langle\varphi\rangle_r \approx 0.1$ and the decay factor in the exponent becomes $\lambda(0.1) - 1 - \ln \lambda(0.1) \approx 2.4$. This cut-off is so strong than not even a trace of a power law distribution remains for such papers.

To compute the overall probability distribution of citations we need to average Eq.(IV.25) over fitness:

$$P(n) \propto \frac{1}{\sqrt{2\pi}n^{3/2}} \int_0^1 \frac{d\varphi}{\lambda(\varphi)} e^{-(\lambda(\varphi)-1-\ln \lambda(\varphi))n} \qquad (IV.26)$$

We will concentrate on the large $n$ asymptotic. Then only highest-fitness papers, which have $\lambda(\varphi)$ close to 1, are important and we can approximate the integral in Eq. (IV.26), using Eq. (IV.8), as:

$$\int_0^1 d\varphi \exp\left(-\left[1-\varphi\frac{1-\alpha}{\langle\varphi\rangle_r}\right]^2 \frac{n}{2}\right) = \frac{\langle\varphi\rangle_r}{1-\alpha}\sqrt{\frac{2}{n}} \int_{\left(1-\frac{1-\alpha}{\langle\varphi\rangle_r}\right)\sqrt{\frac{n}{2}}}^{\sqrt{\frac{n}{2}}} dz\, e^{-z^2}$$

We can replace the upper limit in the above integral with infinity when $n$ is large. The lower limit can be replaced with zero when $n \ll n_c$, where

$$n_c = 2\left(1-\frac{1-\alpha}{\langle\varphi\rangle_r}\right)^{-2} . \qquad (IV.27)$$

In that case the integral is equal to $\sqrt{\pi}/2$, and Eq.(IV.26) gives:

$$P(n) \propto \frac{\langle \varphi \rangle_r}{2(1-\alpha)} \frac{1}{n^2}. \tag{IV.28}$$

In the opposite case, $n \gg n_c$, we get:

$$P(n) \propto \frac{\langle \varphi \rangle_r}{4(1-\alpha)} \frac{\sqrt{n_c}}{n^{2.5}} e^{-\frac{n}{n_c}} \tag{IV.29}$$

When $\alpha = 0.1$, $n_c = 3 \times 10^5$.

Compared to the model without fitness, we have a modified power-law exponent (2 instead of 3/2) and a much relaxed cut-off of this power law. This is consistent with the actual citation data shown in the Fig. IV.2.

As was already mentioned, because of the uncertainty of the definition of "recent" papers, the exact value of $\alpha$ is not known. Therefore, we give $n_c$ for a range of values of $\alpha$ in Table IV.1. As long as $\alpha \leq 0.15$ the value of $n_c$ does not contradict existing citation data.

**Table IV.1.** The onset of exponential cut-off in the distribution of citations, $n_c$, as a function of $\alpha$, computed using Eqs.(IV.27) and (IV24).

| $\alpha$ | 0.3 | 0.25 | 0.2 | 0.15 | 0.1 | 0.05 |
|---|---|---|---|---|---|---|
| $n_c$ | 167 | 409 | 1405 | 9286 | 3.1E+05 | 7.2E+09 |

The major results, obtained for the uniform distribution of fitness, also hold for a non-uniform distribution, which approaches some finite value at its upper extreme $p_p(\varphi = 1) = a > 0$. In Ref. [33] we show that in this case $(1-\alpha)/\langle \varphi \rangle_r$ is very close to unity when $\alpha$ is small. Thus we can treat Eq.(IV.26) the same way as in the case of the uniform distribution of fitness. The only change is that Eqs. (IV.28) and (IV.29) acquire a pre-factor of $a$. Things turn out a bit different when $p_p(1) = 0$. In Appendix C we consider the fitness distribution, which vanishes at $\varphi = 1$ as a power law: $p_p(\varphi) = (\theta + 1)(1 - \varphi)^\theta$.

When $\theta$ is small ($\theta < \frac{2 \times \alpha}{1-\alpha}$) the behavior of the model is similar to what was in the case of a uniform fitness distribution. The distribution of the fitness of cited papers $p_r(\varphi)$ approaches some limiting form with $(1-\alpha)/\langle\varphi\rangle_r$ being very close to unity when $\alpha$ is small. The exponent of the power law is, however, no longer 2 as it was in the case of a uniform fitness distribution (Eq (IV.28)), but becomes $2+\theta$. However, when $\theta > \frac{2 \times \alpha}{1-\alpha}$ the model behaves differently: $(1-\alpha)/\langle\varphi\rangle_r$ strictly equals 1. This means that the power law does not have an exponential cut-off. Thus, a wide class of fitness distributions produces citation distributions very similar to the experimentally observed one. More research is needed to infer the actual distribution of the Darwinian fitness of scientific papers.

The fitness distribution of references $p_r(\varphi)$ adjusts itself in a way that the fittest papers become critical. This is similar to what happens in the Self-Organized Criticality (SOC) model [44] where the distribution of sand grains adjusts itself that the avalanches become critical. Recently we proposed a similar SOC-type model to describe the distribution of links in blogosphere [39].

### B. Distribution of citations to papers cited during the same year

This distribution in the case without fitness is given in Eq. (IV.16). To account for fitness we need to replace $\lambda$ with $\lambda(\varphi)$ in Eq.(IV.16) and integrate it over $\varphi$. The result is:

$$p(n) \sim \frac{1}{n^2} e^{-n/n_c^*}, \qquad \text{(IV.30)}$$

where

$$n_c^* = \frac{1}{2}\left(1 - \frac{1-\alpha}{\langle\varphi\rangle_r}\right)^{-1}. \qquad \text{(IV.31)}$$

Note that $n_c^* \sim \sqrt{n_c}$. This means that the exponential cut-off starts much sooner for the distribution of citation to papers *cited* during the same year, then for citation distribution for papers *published* during the same year.

The above results qualitatively agree with the empirical data for papers cited in 1961 (see Fig.2 in [23]). The exponent of the power law of citation distribution reported in that work is, however, between 2.5 and 3. Quantitative agreement thus may be lacking.

## *4. Effects of literature growth*

Up to now we implicitly assumed that the yearly volume of published scientific literature does not change with time. In reality, however, it grows, and does so exponentially. To account for this, we introduce a Malthusian parameter, $\beta$, which is yearly percentage increase in the yearly number of published papers. From the data on the number of items in the Mathematical Reviews Database [40], we obtain that the literature growth between 1970 and 2000 is consistent with $\beta \approx 0.045$. From the data on the number of source publications in the ISI database (see Table 1 in [30]) we get that the literature growth between 1973 and 1984 is characterized by $\beta \approx 0.03$. One can argue that the growth of the databases reflected not only growth of the volume of scientific literature, but also increase in activities of Mathematical Reviews and ISI and true $\beta$ must be less. One can counter-argue that may be ISI and Mathematical Reviews could not cope with literature growth and $\beta$ must be more. Another issue is that the average number of references in papers also grows. What is important for our modeling is the yearly increase not in number of papers, but in the number of citations these papers contain. Using the ISI data we get that this increase is characterized by $\beta \approx 0.05$. As we are not sure of the precise value of $\beta$, we will be giving quantitative results for a range of its values.

**A. Model without fitness**

At first, we will study the effect of $\beta$ in the model without fitness. Obviously, Equations (IV.2) and (IV.3) will change into:

$$\lambda_0 = \alpha(1+\beta)N_{ref}, \qquad \text{(IV.32)}$$

$$\lambda = (1-\alpha)(1+\beta). \qquad \text{(IV.33)}$$

The estimate of the actual value of $\lambda$ is:

$\lambda \approx (1-0.1)(1+0.05) \approx 0.945$. Substituting this into Equation (IV.9) we get that the exponential cut-off in citation distribution now happens after about 660 citations.

A curious observation is that when the volume of literature grows in time the average amount of citations a paper receives, $N_{cit}$, is bigger than the average amount of references in a paper, $N_{ref}$. Elementary calculation gives:

$$N_{cit} = \sum_{m=0}^{\infty} \lambda_0 \lambda^m = \frac{\lambda_0}{1-\lambda} = \frac{\alpha(1+\beta)N_{ref}}{1-(1-\alpha)(1+\beta)} \tag{IV.34}$$

As we see $N_{cit} = N_{ref}$ only when $\beta = 0$ and $N_{cit} > N_{ref}$ when $\beta > 0$. There is no contradiction here if we consider an infinite network of scientific papers, as one can show using methods of the set theory that there are one-to-many mappings of an infinite set on itself. When we consider real, i.e. finite, network where the number of citations is obviously equal to the number of references we recall that $N_{cit}$, as computed in Eq. (IV.34), is the number of citations accumulated by a paper during its cited lifetime. So recent papers did not yet receive their share of citations and there is no contradiction again.

### B. Model with Darwinian fitness

Taking into account literature growth leads to transformation of Equations (IV.18) and (IV.19) into:

$$\lambda_0(\varphi) = \alpha(1+\beta)N_{ref}\varphi/\langle\varphi\rangle_p, \tag{IV.35}$$

$$\lambda(\varphi) = (1-\alpha)(1+\beta)\varphi/\langle\varphi\rangle_r. \tag{IV.36}$$

As far as the average fitness of a reference, $\langle\varphi\rangle_r$, goes, $\beta$ has no effect. Clearly, its only result is to increase the number of citations to all papers (independent of their fitness) by a factor $1+\beta$. Therefore $\langle\varphi\rangle_r$ is still given by Eq. (IV.23). While, $\lambda(\varphi)$ is always less than unity in the case with no literature growth, it is no longer so when we take this growth into account. *When $\beta$ is large enough, some papers can become supercritical*. The critical value of $\beta$, i.e.

the value which makes papers with $\varphi = 1$ critical, can be obtained from Eq.(IV.36):

$$\beta_c = \langle\varphi\rangle_r /(1-\alpha) - 1. \tag{IV.37}$$

When $\beta > \beta_c$, a finite fraction of papers becomes supercritical. The rate of citing them will increase with time. Note, however, that it will increase always slower than the amount of published literature. Therefore, the relative fraction of citations to those papers to the total number of citations will decrease with time.

Critical values of $\beta$ for several values of $\alpha$ are given in Table IV.2. For realistic values of parameters ($\alpha \leq 0.15$ and $\beta \geq 0.03$) we have $\beta > \beta_c$ and thus our model predicts the existence of supercritical papers. Note, however, that this conclusion also depends on the assumed distribution of fitness.

**Table IV.2** Critical value of the Malthusian parameter $\beta_c$ as a function of $\alpha$ computed using Eq. (IV.37). When $\beta > \beta_c$ the fittest papers become super-critical.

| $\alpha$ | 0.3 | 0.25 | 0.2 | 0.15 | 0.1 | 0.05 |
|---|---|---|---|---|---|---|
| $\beta_c$ | 0.12 | 0.075 | 0.039 | 0.015 | 2.6E-03 | 1.7E-05 |

It is not clear whether supercritical papers exist in reality or are merely a pathological feature of the model. Supercritical papers probably do exist if one generalizes "citation" to include references to a concept, which originated from the paper in question. For instance, these days a negligible fraction of scientific papers which use Euler's Gamma function contain a reference to Euler's original paper. It is very likely that the number of papers mentioning Gamma function is increasing year after year.

Let us now estimate the fraction of supercritical papers predicted by the model. As $(1-\alpha)/\langle\varphi\rangle_r$ is very close to unity, it follows from Eq.(IV.36) that papers with fitness $\varphi > \varphi_c \approx 1/(1+\beta) \approx 1-\beta$ are in the supercritical regime. As $\beta \approx 0.05$, about 5% of papers are in such regime. This does not mean that 5% of papers will be cited forever, because being in supercritical regime only means having extinction probability less than one. To compute this probability we substitute Equations (IV.36) and (IV.5) into Eq.(A3) and get:

$$p_{ext}(\varphi) = \exp((1+\beta) \times \varphi \times (p_{ext}(\varphi) - 1)).$$

It is convenient to rewrite the above equation in terms of survival probability:

$$1 - p_{surv}(\varphi) = \exp(-(1+\beta) \times \varphi \times p_{surv}(\varphi)).$$

As $\beta \ll 1$ the survival probability is small and we can expand the RHS of the above equation in powers of $p_{surv}$. We limit this expansion to terms up to $(p_{surv})^2$ and after solving the resulting equation get:

$$p_{surv}(\varphi) \approx 2 \frac{\varphi - \frac{1}{1+\beta}}{(1+\beta)\varphi} \approx 2(\varphi - 1 + \beta).$$

The fraction of forever-cited papers is thus: $\int_{1-\beta}^{1} 2(\varphi - 1 + \beta) d\varphi = \beta^2$. For $\beta \approx 0.05$ this will be one in four hundred. By changing the fitness distribution $p_p(\varphi)$ from a uniform this fraction can be made much smaller.

## 5. *Numerical simulations*

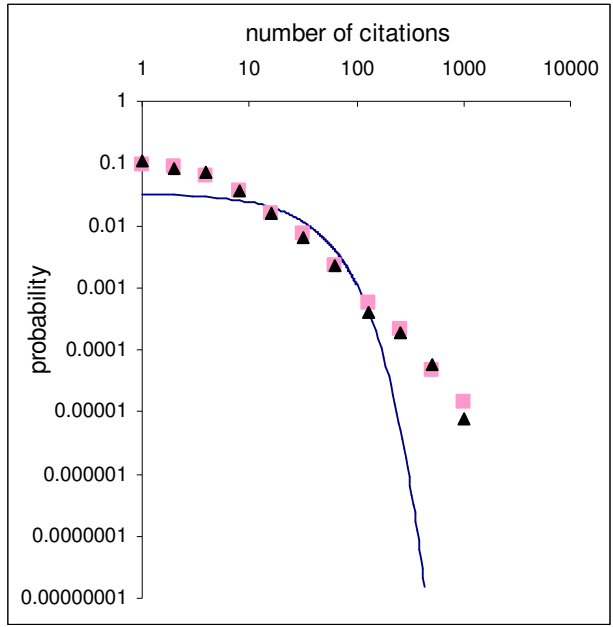

**Figure IV.2.** Numerical simulations of the *modified* model of random-citing scientists (triangles) compared to actual citation data for papers *published during a single year* (squares). The solid line is the prediction of cumulative advantage (AKA preferential attachment) model.

The analytical results are of limited use, as they are exact only for infinitely old papers. To see what happens with finitely old papers, one has to do numerical simulations. Figure IV.2 shows results from such simulations (with $\alpha = 0.1$, $\beta = 0.05$, and uniform between 0 and 1 fitness distribution), i.e., distributions of citations to papers published within a single year, 22 years after publication. Results are compared with actual citation data for Physical Review D papers published in 1975 (as of 1997) [11]. Prediction of the cumulative advantage [20] (AKA preferential attachment [21]) model is also shown. As we mentioned earlier, that model leads to exponential distribution of citations to papers of same age, and thus can not account for highly-skewed distribution empirically observed.

## *6. Aging of scientific literature*

Scientific papers tend to get less frequently cited as time passes since their publication. There are two ways to look at the age distribution of citations. One can take all papers *cited* during a particular year, and study the distribution of their ages. In Bibliometrics this is called *synchronous* distribution [30]. One can take all the papers *published* during a particular distant year, and study the distribution of the citations to these papers with regard to time difference between citation and publication. Synchronous distribution is steeper than the distribution of citation to papers published during the same year (see Figures 2 and 3 in [30]). For example, if one looks at a synchronous distribution, than ten year old papers appear to be cited 3 times less than two year old papers. However, when one looks at the distribution of citations to papers published during the same year the number of citations ten years after publication is only 1.3 times less than two years after publication. The apparent discrepancy is resolved by noting that the number of published scientific papers had grown 2.3 times during eight years. When one plots not total number of citations to papers published in a given year, but the ratio of this number to the annual total of citations than resulting distribution (it is called *diachronous* distribution [30]) is symmetrical to the synchronous distribution.

Recently, Redner [36] who analyzed a century worth of citation data from Physical Review had found that the synchronous distribution (he calls it citations *from*) is exponential, and the distribution of citations to papers published during the same year (he calls it citations *to*) is a power law with an exponent close to one. If one were to construct a diachronous distribution using Redner's data, – it would be a product of a power law and an exponential function. Such distribution is difficult to tell from an exponential one. Thus, Redner's data may be consistent with synchronous and diachronous distributions being symmetric.

The predictions of the mathematical theory of citing are as follows. First, we consider the model without fitness. The average number of citations a paper receives during the $k^{\text{th}}$ year since its publication, $C_k$, is:

$$C_k = \lambda_0 \lambda^{k-1}, \tag{IV.38}$$

and thus, decreases exponentially with time. This is in qualitative agreement with Nakamoto's [30] empirical finding. Note, however, that the exponential decay is empirically observed after the second year, with average number of the second year citations being higher than the first year. This can be understood as a mere consequence of the fact that it takes about a year for a submitted paper to get published.

Let us now investigate the effect of fitness on literature aging. Obviously, Eq. (IV.38) will be replaced with:

$$C_k = \int_0^1 d\varphi \lambda_0(\varphi) \lambda^{k-1}(\varphi). \tag{IV.39}$$

Substituting Eqs.(IV.18) and (IV.19) into Eq.(IV.39) and performing integration we get:

$$C_k = \frac{\alpha N_{ref}}{\langle\varphi\rangle_p} \left(\frac{1-\alpha}{\langle\varphi\rangle_r}\right)^{k-1} \frac{1}{k+1} \tag{IV.40}$$

The average rate of citing decays with paper's age as a power law with an exponential cut-off. This is in agreement with Redner's data (See Fig.7 of [36]), though it contradicts the older work [30], which found exponential decay of citing with time.

In our model, the transition from hyperbolic to exponential distribution occurs after about

$$k_c = -1\big/\ln\!\left((1-\alpha)/\langle\varphi\rangle_r\right) \tag{IV.41}$$

years. The values of $k_c$ for different values of $\alpha$ are given in Table IV.3. The values of $k_c$ for $\alpha \leq 0.2$ do not contradict the data reported by Redner [36].

We have derived literature aging from a realistic model of scientist's referencing behavior. Stochastic models had been used previously to study literature aging, but they were of artificial type. Glänzel and Schoepflin [31] used a modified cumulative advantage model, where the rate of citing is proportional to the product of the number of accumulated citations and some factor, which decays with age. Burrell [41], who modeled citation process as a non-homogeneous Poisson process had to postulate some obsolescence distribution function. In both these cases, aging was inserted by hand. In contrast, in our model, literature ages naturally.

**Table IV.3.** The number of years, after which the decrease in average citing rate will change from a power law to an exponential, $k_c$, computed using Eq.(IV.41), as a function of $\alpha$.

| $\alpha$ | 0.3 | 0.25 | 0.2 | 0.15 | 0.1 | 0.05 |
|---|---|---|---|---|---|---|
| $k_c$ | 9 | 14 | 26 | 68 | 392 | 59861 |

## 7. Sleeping Beauties in science

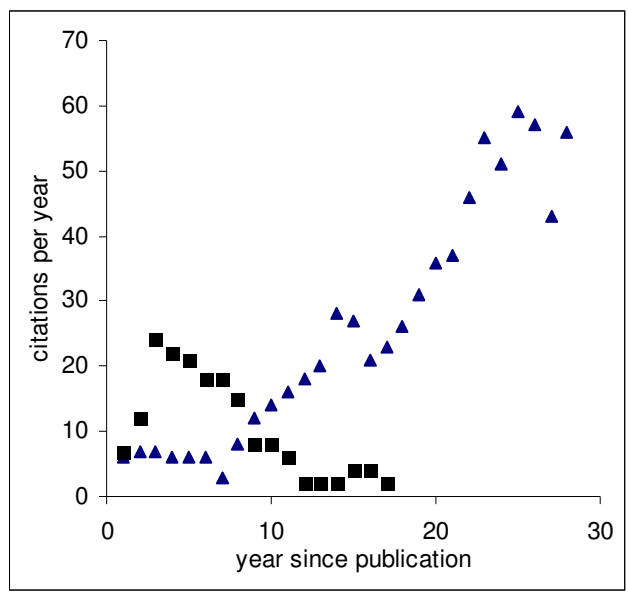

**Figure IV.3.** Two distinct citation histories: an ordinary paper (squares) and a "Sleeping Beauty" (triangles).

Figure IV.3 shows two distinct citation histories. The paper, whose citation history is shown by the squares, is an ordinary paper. It merely followed some trend. When ten years later that trend got out of fashion the paper got forgotten. The paper, whose citation history is depicted by the triangles, reported an important but premature [42] discovery, significance of which was not immediately realized by scientific peers. Only ten years after its publication did the paper get recognition, and got cited widely and increasingly. Such papers are called "*Sleeping Beauties*" [43]. Surely, the reader has realized that both citation histories are merely the outcomes of numerical simulations of the modified model of random-citing scientists.

## *8. Relation to Self Organized Criticality*

Three out of twelve high profile papers misprints in citing which we studied in Section I (see papers 10, 11, and 12 in Tables I.1 and I.2) advance the science of Self Organized Criticality (SOC) [44]. Interestingly this science itself is directly related to the theory of citing. We model scientific citing as a random branching process. In its mean-field version, SOC can also be described as a branching process [45]. Here the sand grains, which are moved during the original toppling, are equivalent to sons. These displaced grains can cause further toppling, resulting in the motion of more grains, which are equivalent to grandsons, and so on. The total number of displaced grains is the size of the avalanche and is equivalent to the total offspring in the case of a branching process. The distribution of offspring sizes is equivalent to the distribution of avalanches in SOC.

Bak [46] himself had emphasized the major role of chance in works of Nature: one sand grain falls, - nothing happens; another one (*identical*) falls, - and causes an avalanche. Applying these ideas to biological evolution, Bak & Sneppen [47] argued that no cataclysmic external event was necessary to cause a mass extinction of dinosaurs. It could have been caused by one of many minor external events. Similarly, in the model of random-citing scientists: one paper goes unnoticed, but another one (*identical in merit*), causes an avalanche of citations. Therefore apart from explanations of $1/f$ noise, avalanches in sandpiles, and extinction of dinosaurs, the highly cited Science of Self Organized Criticality can also account for its own success.

## V. Discussion

The conclusion of this study that a scientific paper can become famous due to ordinary law of chances independently of its content may seem shocking to some people. Here we present more facts to convince them.

Look at the following example. The writings of J. Lacan (10,000 citations) and G. Deleuze (8,000 citations) were exposed by Sokal & Bricmont [49] as nonsense. At the same time, the work of the true scientists is far less cited: A. Sokal – 2,700 citations, J. Bricmont – 1,000 citations.

Additional support for the plausibility of this conclusion gives us the statistics of the very misprints in citations the present study grew from. Few citation slips repeat dozens of times, while most appear just once (see Fig.II.1). Can one misprint be more seminal than the other?

More support comes from the studies of popularity of other elements of culture. A noteworthy case where prominence is reached by pure chance is the statistics of baby-names. Hahn and Bentley [50] observed that their frequency distribution follows a power law, and proposed a copying mechanism that can explain this observation. For example, during the year 2000 34,448 new-born American babies were named Jacob, while only 174 were named Samson [51]. This means that the name "Jacob" is two hundred times more popular than the name "Samson." Is it intrinsically better?

A blind test was administered offering unlabeled paintings, some of which were famous masterpieces of Modern art while others were produced by the author of the test [52]. Results indicate that people cannot tell great art from chaff when the name of a great artist is detached from it. One may wonder if a similar test with famous scientific articles would lead to similar results. In fact there is one forgotten experiment though not with scientific articles, but with a scientific lecture. Naftulin, Ware and Donnelly [53] programmed an actor to teach on a subject he knew nothing. They presented him to a scientific audience as Dr. Myron Fox, an authority on application of mathematics to human behavior (we would like to note that in practice the degree of authority of a scientist is determined by the number of citations to his papers). He read a lecture and answered questions and nobody suspected anything wrong. Afterward the attendees were asked to rate the lecturer and he got high grades. They indicated that they learned a lot from the lecture and one of respondents even indicated that he had read Dr. Fox's articles.

To conclude let us emphasize that the Random-citing model is used not to ridicule the scientists, but because it can be exactly solved using available mathematical methods, while yielding a better match with data than any existing model. This is similar to the random-phase approximation in the theory of an electron gas. Of course, the latter did not arouse as much protest, as the model of random-citing scientists, - but this is only because electrons do not have a voice. What is an electron? - Just a green trace on the screen of an oscilloscope. Meanwhile, within itself, electron is very complex and is as inexhaustible as the universe. When an electron is annihilated in a lepton collider, the whole universe dies with it. And as for the random-phase approximation: Of course, it accounts for the experimental facts - but so does the model of random-citing scientists.

## *Appendix A: Theory of branching processes*

Let us consider a model where in each generation, $p(0)$ per cent of the adult males have no sons, $p(1)$ have one son and so on. The problem is best tackled using the method of generating functions [34], which are defined as:

$$f(z) = \sum_{n=0}^{\infty} p(n) z^n .  \tag{A1}$$

These functions have many useful properties, including that the generating function for the number of grandsons is $f_2(z) = f(f(z))$. To prove this, notice that if we start with two individuals instead of one, and both of them have offspring probabilities described by $f(z)$, their combined offspring has generating function $(f(z))^2$. This can be verified by observing that the *n*th term in the expansion of $(f(z))^2$ is equal to $\sum_{m=0}^{n} p(n-m) p(m)$, which is indeed the probability that the combined offspring of two people is *n*. Similarly one can show that the generating function of combined offspring of *n* people is $(f(z))^n$. The generating function for the number of grandsons is thus:

$$f_2(z) = \sum_{n=0}^{\infty} p(n)(f(z))^n = f(f(z)).$$

In a similar way one can show that the generating function for the number of grand-grandsons is $f_3(z) = f(f_2(z))$ and in general:

$$f_k(z) = f(f_{k-1}(z)).  \tag{A2}$$

The probability of extinction, $p_{ext}$, can be computed using the self-consistency equation:

$$p_{ext} = \sum_{n=0}^{\infty} p(n) p_{ext}^n = f(p_{ext}).  \tag{A3}$$

The fate of families depends on the average number of sons $\lambda = \sum n p(n) = [f'(z)]_{z=1}$. When $\lambda < 1$, Eq. (A3) has only one solution, $p_{ext} = 1$, that is all families get extinct (this is called subcritical branching process). When $\lambda > 1$, there is a solution where $p_{ext} < 1$, and only some of the families get extinct, while others continue to exist forever (this is called supercritical branching process). The intermediate case, $\lambda = 1$, is critical branching proc-ess, where all families get extinct, like in a subcritical process, though some of them only after very long time.

For a subcritical branching process we will also be interested in the probability distribution, $P(n)$, of total offspring, which is the sum of the numbers of sons, grandsons, grand-grandsons and so on (to be precise we include the original individual in this sum just for mathematical convenience). We define the corresponding generating function [54]:

$$g(z) = \sum_{n=1}^{\infty} P(n) z^n \quad . \tag{A4}$$

Using an obvious self-consistency condition (similar to the one in Eq.(A3)) we get:

$$zf(g) = g \tag{A5}$$

We can solve this equation using Lagrange expansion (see Ref. [48]), which is as follows. Let $z = F(g)$ and $F(0) = 0$ where $F'(0) \neq 0$, then:

$$\Phi(g(z)) = \sum_{n=0}^{\infty} \frac{1}{n!} \frac{d^{n-1}}{dg^{n-1}} \left( \Phi'(g) \left( \frac{g}{F(g)} \right)^n \right) \Bigg|_{g=0} z^n \tag{A6}$$

Substituting $F(g) = g/F(g)$ (see Eq. (A5)) and $\Phi(g) = g$ into Eq. (A6) we get:

$$g = \sum_{n=1}^{\infty} \frac{z^n}{n!} \left[ \frac{d^{n-1}}{d\omega^{n-1}} (f(\omega))^n \right]_{\omega=0} \tag{A7}$$

Using Eq.(A4) we get:

$$P(n) = \frac{1}{n!} \left[ \frac{d^{n-1}}{d\omega^{n-1}} (f(\omega))^n \right]_{\omega=0} \tag{A8}$$

Theory of branching processes can help to understand scientific citation process. The first year citations correspond to sons. Second year citations, which are copies of the first year citations, correspond to grandsons, and so on.

## *Appendix B*

Let us consider the case when $\lambda \neq \lambda_0$, i.e. a branching process were the generating function for the first generation is different from the one for subsequent generations. One can show that the generating function for the total offspring is:

$$\tilde{g}(z) = z f_0(g(z)).  \quad (B1)$$

In the case $\lambda = \lambda_0$ we have $f(z) = f_0(z)$ and because of Eq. (A5) $\tilde{g}(z) = g(z)$. We can compute $f_0(g(z))$ by substituting $f_0$ for $\Phi$ in Eq.(A6)

$$f_0(g(z)) = \sum_{n=0}^{\infty} \frac{1}{n!} \frac{d^{n-1}}{dg^{n-1}} \left( f_0'(g)(f(g))^n \right) \bigg|_{g=0} z^n  \quad (B2)$$

After substituting Eqs. (IV.4) and (IV.5) into Eq. (B2) and the result into Eq. (B1) we get

$$\tilde{P}(n) = \frac{\lambda_0 ((n-1)\lambda + \lambda_0)^{n-2}}{(n-1)!} e^{-(n-1)\lambda - \lambda_0}.  \quad (B3)$$

The large *n* asymptotic of Eq. (B3) is

$$\tilde{P}(n) \propto \frac{\lambda_0}{\lambda} \exp\left( \frac{\lambda_0}{\lambda} - 1 + \lambda - \lambda_0 \right) P(n)  \quad (B4)$$

where $P(n)$ is given by Eq.(IV.7). We see that having different first generation offspring probabilities does not change the functional form of the large-*n* asymptotic, but merely modifies the numerical prefactor. After substituting $\alpha \approx 0.1$ and $N_{ref} \approx 20$ into Eqs. (IV.2) and (IV.3) and the result into Eq.(B4) we get $\tilde{P}(n) \approx 2.3 P(n)$.

## *Appendix C*

Let us investigate the fitness distribution

$$p_p(\varphi) = (\theta+1)(1-\varphi)^\theta \tag{C1}$$

After substituting Eq. (C1) into Eq. (IV.21) we get:

$$p_r(\varphi) = \frac{\alpha(\theta+1)(\theta+2)\varphi(1-\varphi)^\theta}{1-((1-\alpha)/\langle\varphi\rangle_r)\varphi} \tag{C2}$$

After substituting this into Eq. (IV.22) we get:

$$1 = \alpha(\theta+1)(\theta+2)\left(\frac{\langle\varphi\rangle_r}{1-\alpha}\right)^2 \int_0^1 \frac{(1-\varphi)^\theta\, dx}{\frac{\langle\varphi\rangle_r}{1-\alpha}+\varphi} - \alpha(\theta+2)\frac{\langle\varphi\rangle_r}{1-\alpha} \tag{C3}$$

As acceptable values of $\langle\varphi\rangle_r$ are limited to the interval between $1-\alpha$ and 1, it is clear that when $\alpha$ is small the equality in (C3) can only be attained when the integral is large. This requires $\frac{\langle\varphi\rangle_r}{1-\alpha}$ being close to 1. And this will only help if $\theta$ is small. In such case the integral in Eq.(C3) can be approximated as

$$\int_{\frac{\langle\varphi\rangle_r}{1-\alpha}}^1 \frac{\varphi^\theta\, dx}{\varphi} = \frac{1}{\theta}\left(1-\left(\frac{\langle\varphi\rangle_r}{1-\alpha}-1\right)^\theta\right).$$

Substituting this into Eq.(C3) and replacing in the rest of it $\frac{\langle\varphi\rangle_r}{1-\alpha}$ with unity we can solve the resulting equation to get:

$$\frac{\langle\varphi\rangle_r}{1-\alpha}-1 \approx \left(\frac{\alpha-\dfrac{\theta}{\theta+2}}{\alpha(\theta+1)}\right)^{\frac{1}{\theta}} \tag{C4}$$

For example, when $\alpha = 0.1$ and $\theta = 0.1$ we get from Eq.(C4) that $\dfrac{\langle\varphi\rangle_r}{1-\alpha}-1 \approx 6\times 10^{-4}$. However Eq.(C4) gives a real solution only when

$$\alpha \geq \frac{\theta}{\theta+2}. \tag{C5}$$

The R.H.S. of Eq.(C2) has a maximum for all values of $\varphi$ when $\langle\varphi\rangle_r = 1-\alpha$. After substituting this into Eq.(C2) and integrating we get that the maximum possible value of $\int_0^1 p_r(\varphi)d\varphi$ is $\alpha\dfrac{\theta+2}{\theta}$. We again get a problem when the condition of Eq.(C5) is violated. Remember, however, that when we derived Eq.(IV.21) from Eq.(IV.20) we divided by $1-(1-\alpha)\varphi/\langle\varphi\rangle_r$, which, in the case $\langle\varphi\rangle_r = 1-\alpha$, is zero for $\varphi = 1$. Thus, Eq.(IV.21) is correct for all values of $\varphi$, except for 1. The solution of Eq.(IV.20) in the case when the condition of Eq.(C5) is violated is:

$$p_r(\varphi) = \alpha(\theta+1)(\theta+2)\varphi(1-\varphi)^{\theta-1} + \left(1-\alpha\frac{\theta+2}{\theta}\right)\delta(\varphi-1) \tag{C6}$$